\documentclass[aps,twocolumn,prl,10pt]{revtex4-2}

\usepackage{color} 
\usepackage{amsmath}
\usepackage{amssymb}
\usepackage{tabularx}
\usepackage{multirow}
\usepackage{makecell}
\usepackage{graphicx}
\usepackage{physics}
\usepackage[dvipsnames]{xcolor}

\usepackage[left, mathlines]{lineno}

\usepackage{url} 

\newcommand{\mos}{\ensuremath{\mathrm{MoS}_2}}
\newcommand{\wse}{\ensuremath{\mathrm{WSe}_2}}
\newcommand{\hfs}{{\mathrm{HfS}_2}}
\newcommand{\gas}{{\mathrm{Ga}\mathrm{S}}}
\newcommand{\moire}{moir\'e }
\newcommand{\Moire}{Moir\'e }

\newcommand{\ionR}{\mathbf{R}}
\newcommand{\kvec}{\mathbf{k}}

\newcommand{\neighbours}{\mathcal{N}}

\newcommand{\ilspacing}{d}
\newcommand{\Eintra}{E_{\mathrm{intra}}}
\newcommand{\Einter}{E_{\mathrm{inter}}}

\begin{document}

\title{Accurate, transferable, and verifiable machine-learned interatomic potentials for layered materials}

\author{Johnathan D. Georgaras}
\thanks{These authors contributed equally}
\affiliation{Department of Materials Science and Engineering, Stanford University, Stanford, CA 94305, USA}

\author{Akash Ramdas}
\thanks{These authors contributed equally}
\affiliation{Department of Materials Science and Engineering, Stanford University, Stanford, CA 94305, USA}

\author{Chung Hsuan Shan}
\affiliation{Department of Materials Science and Engineering, Stanford University, Stanford, CA 94305, USA}

\author{Elena Halsted}
\affiliation{Department of Materials Science and Engineering, Stanford University, Stanford, CA 94305, USA}

\author{Berwyn}
\affiliation{Department of Materials Science and Engineering, Stanford University, Stanford, CA 94305, USA}

\author{Tianshu Li}
\affiliation{Department of Materials Science and Engineering, Stanford University, Stanford, CA 94305, USA}

\author{Felipe H. da Jornada}
\email{jornada@stanford.edu}
\affiliation{Department of Materials Science and Engineering, Stanford University, Stanford, CA 94305, USA}

\begin{abstract}
Twisted layered van-der-Waals materials often exhibit unique electronic and optical properties absent in their non-twisted counterparts.
Unfortunately, predicting such properties is hindered by the difficulty in determining the atomic structure in materials displaying large \moire domains.
Here, we introduce a split machine-learned interatomic potential and dataset curation approach that separates intralayer and interlayer interactions and significantly improves model accuracy -- with a \emph{tenfold} increase in energy and force prediction accuracy relative to conventional models.
We further demonstrate that traditional MLIP validation metrics -- force and energy errors -- are inadequate for \moire structures and develop a more holistic, physically-motivated metric based on the distribution of stacking configurations. This metric effectively compares the entirety of \emph{large-scale} \moire domains between two structures instead of relying on conventional measures evaluated on smaller commensurate cells.
Finally, we establish that one-dimensional instead of two-dimensional \moire structures can serve as efficient surrogate systems for validating MLIPs, allowing for a practical model validation protocol against explicit DFT calculations.
Applying our framework to $\hfs/\gas$ bilayers reveals that accurate structural predictions directly translate into reliable electronic properties. 
Our model-agnostic approach integrates seamlessly with various intralayer and interlayer interaction models, enabling computationally tractable relaxation of \moire materials, from bilayer to complex multilayers, with rigorously validated accuracy.

\end{abstract}

\maketitle

\section{Introduction}
\label{sec:intro}

Stacking individual layers of two-dimensional (2D) materials with small misalignments in rotation, strain, or lattice geometry form \textit{moiré superlattices}: structures characterized by periodic unit-cell repetitions that produce long-range modulation patterns spanning tens to hundreds of nanometers. These superlattices host spatially varying stacking-dependent potentials, often leading to flat electronic bands that drive correlated macroscopic phenomena such as unconventional superconductivity~\cite{cao2018unconventional, yankowitz2019tuning, xia2024superconductivity}, exotic topological phases~\cite{tong2017topological,sharpe2019emergent, serlin2020intrinsic}, and correlated insulating states~\cite{cao2018correlated, wu2019topological,chen2020tunable,kerelsky2019maximized}. Moiré patterns also host unique quasiparticles and collective excitations such as moiré-confined excitons, phonons, and spatially trapped states~\cite{seyler2019signatures,lin2018moire,tran2019evidence, karni2022structure}. Predicting and controlling these emergent phenomena necessitates models capable of capturing the strain-induced lattice distortions and distribution of stacking configurations intrinsic to moiré materials.

The conventional approach for modelling \moire structures typically employs density-functional theory (DFT), which can accurately predict atomistic reconstructions and domain formation~\cite{fang2015ab, carr2020electronic}. However, the high computational cost of these methods severely restricts their application to supercells containing only a few thousand atoms~\cite{naik2018ultraflatbands}. Still, these atomistic methods, performed on smaller commensurate lattices, can nonetheless parametrize faster classical and (semi)empirical methods applicable to larger superlattices~\cite{naik2022twister, naik2019kolmogorov}. 

A complementary approach, based on continuum elastic theory, integrates out atomic-scale details to operate on larger-scale displacement fields~\cite{carr2020electronic, carr2018relaxation, cazeaux2020energy,cazeaux2023relaxation,engelke2023topological}.
While not strictly atomistic, such elastic-theory methods can successfully predict and interpret several moiré-related phenomena~\cite{carr_electronic-structure_2020} and clarify the critical role that local stacking energies and strain distributions play in these systems.

A promising generalizable and scalable way to model materials is with machine-learned interatomic potentials (MLIPs). MLIPs combine the computational efficiency of classical force fields with near \emph{ab initio} accuracy through data-driven training on extensive DFT calculation datasets ~\cite{jacobs2025practical}. Most MLIPs generally span a spectrum of interpretability. Kernel-based approaches, like Gaussian approximation potentials~\cite{bartok2010gaussian}, use structured, interpretable descriptors -- \textit{e.g.}, the smooth overlap of atomic positions~\cite{de2016comparing}, atom-centered symmetry functions~\cite{behler2011atom}, or many-body tensor representation~\cite{huo2022unified} -- to interpolate potential energy surfaces with systematic error control and excellent transferability~\cite{musil2021physics, bartok2013representing}. Conversely, neural network potentials leverage their flexibility as universal function approximators to model high-dimensional, non-linear potential energy surfaces.
Recent developments in E(3)-equivariant and message-passing schema~\cite{batzner2023, thomas2018tensor} explicitly encode Euclidean symmetries, enabling physically correct transformations of scalar, vector, and tensor quantities while requiring significantly fewer DFT calculations for model training~\cite{batzner2023}.

Despite these architectural advancements, generating accurate models for \moire materials in practice is still nontrivial~\cite{deringer2020general, magorrian2024strong, nair2024machine}. Suitable MLIPs for \moire materials must precisely capture the balance between strong intralayer covalent bonds  -- about 1 eV per formula unit (f.u.) --  and weak interlayer van der Waals (vdW) interactions -- about 100 meV/f.u. Beyond the training challenge posed by these disparate energy scales, the validation of such models is equally critical. Conventional validation metrics, such as a model energy and force root-mean-square error (RMSE), inadequately assess \moire relaxation quality as they primarily evaluate local atomic environments while relevant \moire structural effects predominantly emerge at length scales beyond those captured by local descriptors. Additionally, the practical limitations of DFT calculations prevent explicit validation of MLIPs on realistic \moire structures.

Here, we introduce a \textit{split} MLIP approach tailored for multilayer moiré systems to address these limitations. By separately modelling the intralayer and interlayer interactions, our method improves accuracy tenfold for force and energy compared to traditional methods, which we refer to as a \textit{unified} model. Our approach remains agnostic to the specific MLIP architecture used for the interlayer and intralayer components and accurately predicts complex \moire structures across various twist angles and number of layers.

Our work also addresses the validation of MLIPs for \moire materials.
We demonstrate the limitations of traditional metrics -- energy and force RMSE -- in assessing the quality of an MLIP, and introduce a metric based on the distribution of stacking registries as a robust alternative.
We further show that MLIPs can be explicitly validated against ground-truth calculations on quasi-1D systems, which exhibit large domain formation similar to their 2D counterparts while remaining accessible to DFT calculations. Importantly, this correlation ensures that an MLIP's fitness in describing 1D structures reliably predicts its fitness in describing 2D \moire systems.

As a case study, we apply our framework to bilayer HfS$_2$/GaS. We use quasi-1D structures to validate our split MLIP structural relaxations against DFT calculations and highlight the similarity between the electronic properties obtained at these two geometries. For 2D \moire HfS$_2$/GaS, we predict strong relaxation effects, with triangular domain formation that could support unusual low-energy electronic states.

Our split MLIP strategy, combined with a rigorous surrogate-based validation protocol, enables accurate and efficient exploration of structural and emergent phenomena in previously inaccessible multilayer moiré systems beyond TMDs.

We present our results in four parts: (1) we detail our split MLIP procedure and demonstrate its improvement over standard approaches; (2) we introduce the mean disregistry error (MDE) metric and establish its relationship with RMSE; (3) we demonstrate how quasi-1D-periodic structures effectively capture the accuracy of 2D moiré reconstructions; and (4) we apply this framework to bilayer HfS$_2$/GaS.

\section{Results}

\subsection{Need for split MLIPs for layered materials}

\begin{figure}[h!]
    \centering
    \includegraphics[width=0.9\columnwidth]{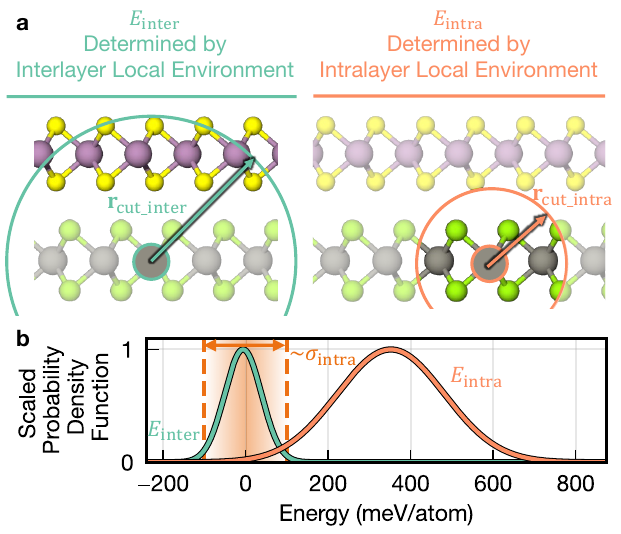} 
    \caption{Separation of length scales for designing accurate MLIPs for layered materials. \textbf{a} Total energy for a layered system can be separated into an intralayer term involving non-interacting layers and a residual interlayer interaction, which naively requires models with longer truncation distances. \textbf{b} Distributions of intralayer and interlayer energies within our generated datasets. The distribution of intralayer energies resembles that in the training of MLIPs with standard approaches, but typical interlayer energies are much smaller -- on the order of the \emph{error} $\sigma_\mathrm{intra}$ of the intralayer MLIP model ($\sim$100 meV/atom) --, motivating the split model approach.
    }
    \label{fig:SplitApproach}
\end{figure}

We first motivate and introduce our proposed methodology for the generation of MLIPs for \moire reconstructions of layered materials. We begin by expressing the energy $E$ of a material with $N$ atoms at positions $\{\ionR_i , i \in {1, 2, ..., N}\}$ in terms of tensor contractions, including correlations of atomic positions of increasingly higher order,
\begin{equation}
    \begin{split}
        E &= \sum_i V^{(i)}_{1} + \sum_{\substack{i \\ j_1 \in \neighbours(i) } } V^{(i,j_1)}_2 (\ionR_{ij_1}) \\
        & +\sum_{\substack{i \\ j_1, j_2 \in \neighbours(i) } } V^{(i,j_1, j_2)}_3 (\ionR_{ij_1}, \ionR_{ij_2}) + ..., 
\end{split}
\end{equation}

where $\neighbours(i)$ is the bonding neighbourhood of atom $i$, $V^{(i,j_1, j_2, ..., j_m)}_{m + 1}$ is the $m + 1$ body potential between atoms $i, j_1, j_2, ..., j_m$, and $\ionR_{ij}$ are the relative positions of the atoms ($\ionR_{ij} = \ionR_{i} - \ionR_{j}$). This form is used explicitly within most MLIPs, and implicitly even in MLIPs utilizing message-passing~\cite{batzner2023}.

An important insight from this paper is that when dealing with layered materials, a similar partition can be formally performed at the layer level. Namely, one can partition the total energy of a layered material from the intralayer energies of atoms in isolated layers plus an interlayer energy contribution from the correlation of atomic positions within two, three, and arbitrarily large numbers of layers. This separation is motivated by the fact that stronger covalent bonds dominate the energy within a single layer, while weaker vdW interactions are responsible, in this framework, for the residual coupling between different layers. 

Current state-of-the-art MLIPs typically construct bonding neighbouring lists $\neighbours(i)$ by including atoms within some cutoff radius $\mathbf{r}_{\mathrm{cut}}$ from atom $i$. Hence, a practical way to split the total energy into intralayer and interlayer contributions is to define different cutoff radii when constructing these two energy contributions (Fig.~1a) and modify the neighbour list to capture these two types of interactions (see SI~\cite{SI}). For this work, we choose  $\mathbf{r}_{\mathrm{cut\_intra}} = 6.0$Å and $\mathbf{r}_{\mathrm{cut}} = \mathbf{r}_{\mathrm{cut\_inter}} = 10.0$Å. 

Figure~1b displays the distribution of intralayer and interlayer energies in a typical moiré system, bilayer $\mos/\mos$, obtained by fitting dataset energies, detailed in the next section, to Gaussian distributions. Notably, the energy scales associated with interlayer interactions are much smaller than the intralayer energies -- and of the order of the \emph{error} of typical MLIPs fit to DFT calculations (shaded region in Fig.~1b). This makes it hard to describe intralayer and interlayer interactions from standard approaches accurately, \textit{e.g.}, training unified MLIPs from molecular dynamics simulation, which we explicitly quantify later. In the next section, we address how to efficiently generate datasets and train different MLIPs for the intralayer and interlayer contributions.

\subsection{Dataset generation protocol for split MLIPs} \label{sec:alldset}

A key aspect in generating MLIPs is that the training dataset must sample a diverse set of atomic configurations that capture the bonding environment relevant for subsequent evaluations of the MLIP. A standard approach when training MLIPs is to sample configurations from molecular dynamics trajectories obtained from ab initio or surrogate models. While such an approach could directly be used in training intralayer MLIPs, we chose a more straightforward procedure here, in which we apply random perturbations to the atomic positions and cell parameters starting from a suitable supercell ($3\times3$ for the calculations in this paper). The advantage of this procedure is that this dataset can be simply generated from parallel DFT calculations and minimizes the chance of sampling unphysical trajectories that are more likely with two-dimensional materials. We then predict the energies and forces of these structures using DFT with a suitable vdW-corrected functional~\cite{cooper2010van} to ensure the intralayer and interlayer datasets were generated with similar DFT parameters (see Methods).

For the training of interlayer MLIPs for a particular pair of layered materials, we generate a dataset that samples various interlayer distances and local stacking environments. In particular, we first construct small unit cells or supercells appropriate to hold the relevant bilayer structure. This work focuses on relaxations near 0° twist; hence, we consider a bilayer at a 0° twist angle. Our model for the bilayer at 0° is quite accurate on the configurations at 60° as well (interlayer force RMSE: 32.86 meV/Å, interlayer energy RMSE: 6.32 meV/atom), making it generalizable to twist angles quite far away from 0° as well. Of course, one can improve this accuracy by adding the 60°-twisted structures into the dataset.

Still considering the interlayer dataset, we sample various layer separations $d$, more so finely near equilibrium, and coarsely otherwise(see Methods). We also sample a set of $12 \times 12$ stacking configurations, out of which only a 6$\times$6 subset is kept in our training and validation sets, with the remaining forming our test set. When computing the interlayer energy and forces for a particular bilayer configuration, we subtract the intralayer contributions from each separate layer. We did not randomize the atomic positions within each layer when constructing the interlayer dataset.
 
For benchmark purposes, we also consider unified MLIP models not separated into intralayer and interlayer parts. For these models, we create a dataset formed by the union of the intralayer and interlayer datasets.
 
\subsection{Split MLIPs benchmark: tenfold improvement of energy and force RMSE metrics} \label{sec:mlipperf}

\begin{figure}[t]
    \centering
    \includegraphics[width=0.9\columnwidth]{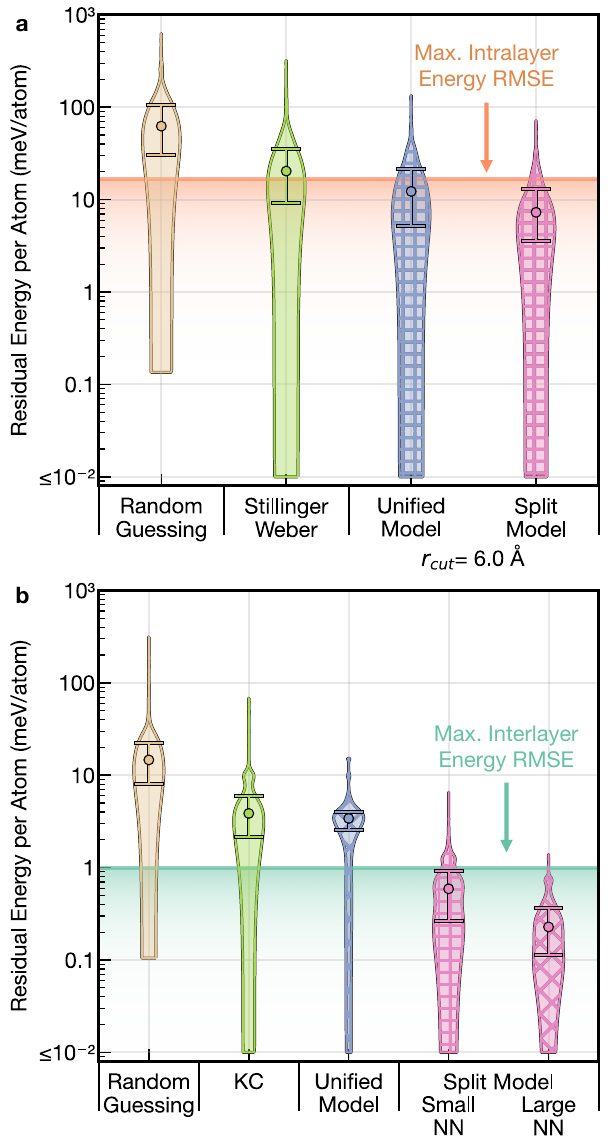}
    \caption{Performance of classical interatomic potentials and both unified and split MLIPs on our hold-out test datasets. \textbf{a} Absolute residual energy, relative to DFT, for various intralayer models. Error bars mark the interquartile range, and dots mark the median absolute residuals. 
    Horizontal lines are the maximum desired model errors (see text). Split and unified models perform similarly on the intralayer predictions when using similar model parameters. \textbf{b} Similar comparison for interlayer models.  Here, even small split MLIPs (6080 trainable parameters) outperform large unified MLIPs (109824 trainable parameters).
    }
    \label{fig:modelperf}

\end{figure}

We now present the first clear benefits of our proposed split MLIP approach for a prototypical bilayer of $\mos$/$\wse$. We train intralayer ($\wse$, $\mos$) and an interlayer ($\mos$/$\wse$) MLIPs to the split datasets using the Allegro model~\cite{musaelian2023learning}, though we find similar results with other MLIPs (SI Fig.~S1~\cite{SI}). 
In Fig.~2a, we compare the RMSEs evaluated on our hold-out $\wse$ intralayer test datasets for several models: a random, mean-guessing baseline, the classical modified Stillinger Weber~\cite{jiang2019misfit}, a unified MLIP with a cutoff of 6 Å, and the split intralayer MLIP with a cutoff of 6 Å. All MLIPs outperform the classical SW potential and, notably, similar split and unified MLIPs (\textit{i.e.}, constructed with a similar number of trainable parameters) display a similar accuracy on the intralayer dataset. This is expected, as the total energy of a unified MLIP is dominated by the intralayer contribution (Fig.~1b).
The accuracy of both unified and split MLIPs in describing intralayer interactions can be improved by making the cutoff radius of the model larger (SI Fig.~S1a~\cite{SI}).

Next, we benchmark the ability of several models to capture interlayer interactions. In Fig.~1b, we evaluate the RMSEs on our hold-out interlayer test datasets for a random guessing, classical Kolmogorov-Crespi (KC), unified MLIP with 109824 trainable parameters, and two split MLIPs with 6080 and 109824 parameters. Notably, the performance of the unified MLIP model is comparable to the classical KC model. However, the split model, even with significantly fewer model parameters, outperforms the unified MLIP. The lower error in these split MLIPs is required for achieving an energy RMSE below a threshold of 1~meV/atom associated with accurate moiré relaxations, which we discuss in the following section.

Our results for the split models are not unique to the Allegro MLIP architecture tested here, and we observe similar trends when employing the MACE~\cite{batatia2022mace} potential (SI Fig.~S1~\cite{SI}). In fact, one could even use different model architectures for each interaction, \textit{e.g.}, a foundation model~\cite{batatia2023foundation, deng2023chgnet} for the intralayer and a system-specific classical potential~\cite{nielsen2023accurate} for the interlayer model. This makes the splitting procedure and dataset generation \emph{model-agnostic}, allowing models to improve at the rapid pace of MLIP improvements.

This approach also simplifies the training and validation of high-quality intralayer MLIP models, given that the number of required intralayer MLIPs grows linearly with the number of layers one is interested in simulating, but the number of pairwise interlayer MLIPs, quadratically. We also note that we obtain the same conclusions when analyzing the force RMSE across these models (SI Fig.~S1~\cite{SI}).

While we have demonstrated that splitting the MLIPs achieves better energy and force RMSE on the test datasets, our ultimate goal is to accurately determine the structure of \moire materials. We next show that the energy and force RMSE are, unfortunately, not good predictors of a model's ability to describe \moire relaxations. For that, we introduce a metric to assess the quality of the structural relaxation. This allows one to determine the maximum permissible energy and force RMSEs (solid lines in Fig.~2), and create a general approach to designing accurate classical and machine-learned IPs for \moire structure relaxation.

\label{sec:vordisreg}

\subsection{Limitations of energy and force errors as validation metrics}

\begin{figure*}[t!]
    \centering
    \includegraphics[width=0.8\linewidth]{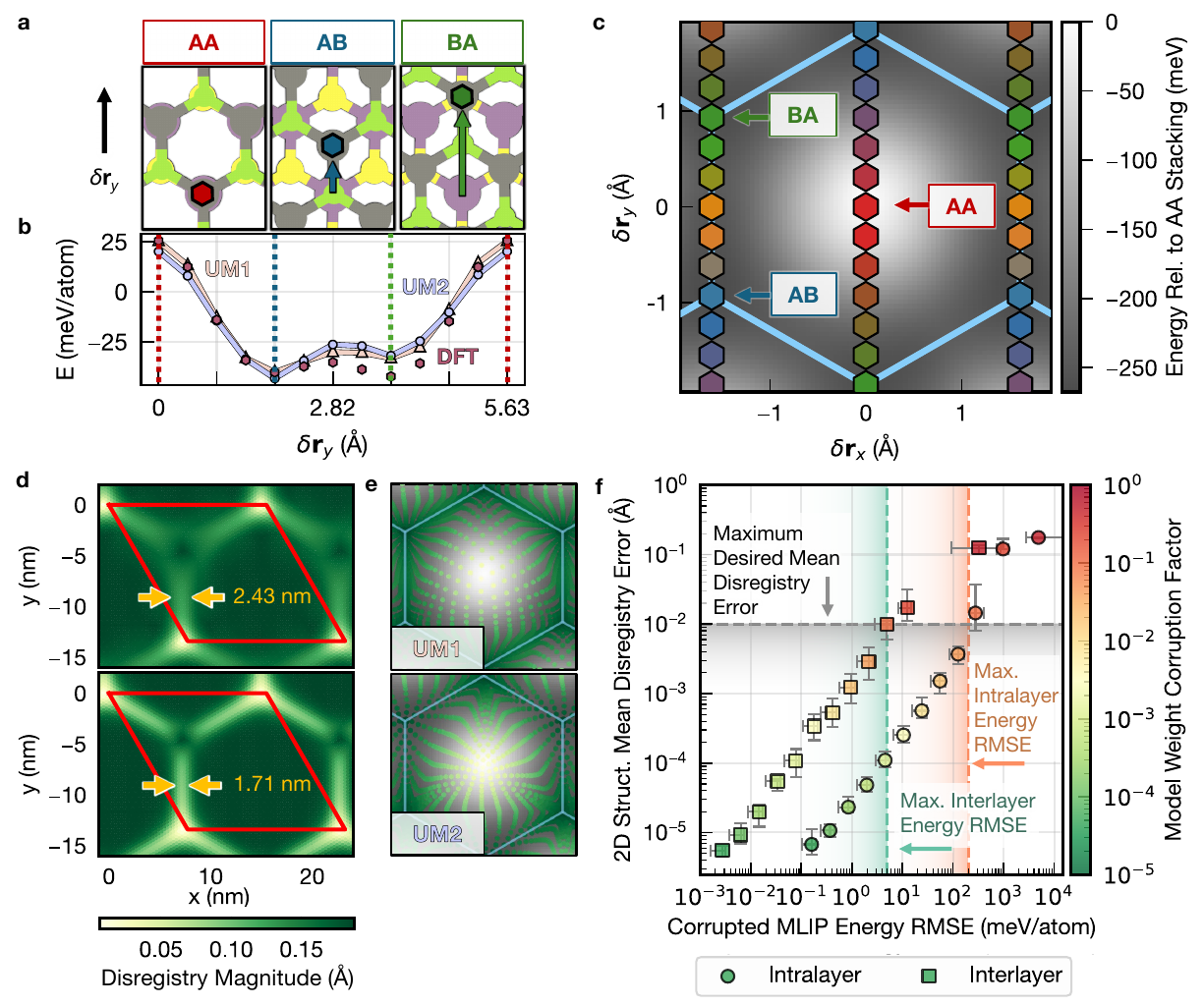}
    \caption{Validation metrics for \moire structural effects in twisted $\mos$/$\wse$ heterostructures. \textbf{a} Atomic configurations at high-symmetry stacking positions (AA, AB, and BA), showing W atoms (coloured hexagons), reference Mo atoms, and disregistry vectors (arrows). \textbf{b} Interlayer potential energy landscape per atom versus displacement ($\delta r_y$) for a ground-truth density functional theory (DFT) prediction and two machine-learned interatomic potential (MLIP) models trained on a unified dataset:  Unified Model 1 (UM1) with a neighbour cutoff radius of 6Å, and Unified Model Biased 2 (UM1) with a neighbour cutoff radius of 10Å. Vertical lines indicate high-symmetry stacking configurations: AA (red), AB (blue), and BA (green). \textbf{c} Voronoi-centered disregistry vectors during continuous displacement of top $\wse$ layer, mapped onto the pristine $\mos$ lattice (blue background cell). \textbf{d} Atomic relaxation patterns in a 1.1$^{\circ}$ twisted $\mos$/$\wse$ heterostructure predicted using UM1 and UM2. The colour scale indicates local disregistry magnitude at Mo sites, revealing domain formation through regions of constant disregistry. \textbf{e} Distribution of Voronoi-centered disregistry vectors of the same twisted structures in 3d, highlighting domain asymmetry through the density of vector clustering near high-symmetry positions. \textbf{f} Structural validation of \moire geometry using the mean disregistry error (MDE) metric versus model energy prediction RMSE for MLIPs trained with varying degrees of model corruption. Shaded regions denote the desired target MDE (grey, below $0.01$ Å), and the maximum interlayer energy RMSE (green, $\leq 0.4$ meV/atom) and maximum intralayer energy RMSE (orange, $\leq200$ meV/atom) to achieve that target MDE.}
    \label{fig:RMSE}
\end{figure*}

MLIPs are typically trained by minimizing a loss function based on the energy and force RMSE relative to ground-truth values, often from DFT calculations~\cite{Behler2007, Bartok2010}. This training process relies on feature vectors in the MLIPs that encode the \emph{local} atomic environment (\textit{e.g.}, Eq.~1)~\cite{zuo2020performance,bartok2013representing, Bartok2010, behler2016perspective}. So, while the energy and force RMSE is a convenient optimization target for most MLIPs, it does not directly assess the relaxation of moiré materials, which critically takes place at longer, nanometer scales.

We illustrate these limitations using $\mos$/$\wse$ bilayers in Fig.~3. In these heterostructures, small energy asymmetries ($\sim$2 meV/atom difference) between the AB and BA stacking configurations fundamentally determine moiré-scale domain formation, domain walls, and resultant electronic properties~\cite{naik2018ultraflatbands}.

To test the model validity of standard RMSE metrics, we analyzed a 0°-twisted $\mos$/$\wse$ bilayer with lattice-matched primitive cells. We evaluated total energy across a systematic displacement path of the top W layer relative to the bottom Mo layer (Fig.~3a). This displacement generates a path in the configuration space, which encodes the relative offset between corresponding atoms in adjacent layers (Fig.~3a). We then evaluated the total energy for this set of structures using two similar MLIP models trained on identical unified datasets but with different cutoff radii: 6~Å for the unified model 1 (UM1) and 10~Å for the unified model 2 (UM2). Both models achieved low energy RMSE values relative to the DFT ground truth (11.6 meV/atom for UM1 and 10.6 meV/atom for UM2), well below the typical energy RMSE of 100 meV/atom target for many MLIP training protocols. 

Despite their low RMSE values, both models incorrectly predict AB stacking as more energetically favourable than BA by about 7 and 11 meV/atom for UM1 and UM2, respectively, which has quantifiable consequences for the models' abilities to predict the structure of \moire systems.
Clearly, a preferred approach for validating MLIPs is to compare the relaxed \moire geometries directly. In principle, traditional MLIP descriptors could be utilized for this, as they often display desired properties such as being invariant under global rotations, translations, and relabeling of atomic species. However, common descriptors typically lack simple physical interpretability, \textit{e.g.}, how their various components translate into the similarity between \moire structures in real space.

To directly obtain a physically interpretable comparison metric, we compare \moire structures at a coarser level in terms of \emph{disregistry vectors}, which encode local stacking environments in terms of a real-space displacement vector $\delta\mathbf{r}$ of one layer relative to a reference high-symmetry stacking configuration (Fig.~3a). For our calculations on $\mos$/$\wse$, we define $\delta\mathbf{r}{=}0$ when the W and Mo atoms are on top of each other (\textit{i.e.}, the AA stacking for a 0°-twisted structure). The vector space associated with the disregistry vectors is the \textit{configuration space}.

Disregistry vectors are not uniquely defined and often expressed within effective (strained) primitive cells -- constructed following a consistent global order to tile any given bilayer material~\cite{massatt2017electronic, carr2017twistronics, carr2018relaxation, cazeaux2020energy}.
We avoid this requirement in our work by instead introducingVoronoi-centered disregistry vectors. In short, we define $\delta\mathbf{r}$ to have the smallest magnitude, which uniquely defines them and increases their applicability to \moire materials with larger strain inhomogeneities(see SI~\cite{SI}).

To illustrate the Voronoi-centered disregistry vector approach, we map the disregistry path in Fig.~3b onto a denser path in configuration space in Fig. 3c, where each hexagon at $\delta\mathbf{r}$ represents a structure wherein the W atom on one layer is laterally displaced from the nearest Mo atom in the adjacent layer by $\delta\mathbf{r}$.

When applied to an explicit moiré superlattice, this mapping transforms collections of local atomic stacking environments into a comprehensive fingerprint of the moiré geometry~\cite{musil2021physics}. As shown in Fig. 3e, relaxed bilayer structures exhibit clustering of disregistry vectors around energetically favourable stacking configurations, with the density of vectors proportional to the corresponding domain area in real space. The spatial distribution of these vectors directly reflects the balance between interlayer stacking energy minimization and elastic energy penalties from deformation and enables quantitative comparison of structural predictions across different models.

We now revisit the differences between UM1 and UM2 by analyzing their relaxations in terms of their disregistry vectors.
We first plot the relaxed structure for a 1.1$^\circ$ twisted $\mos$/$\wse$ predicted by these two MLIPs in Fig.~3d, showing the magnitude $|\delta\mathbf{r}|$ for each atomic position in the \moire cell. UM2, which significantly (and incorrectly) favours AB stacking, produces larger AB moiré domains compared to UM1. Critically, even though both models have similar and relatively small energy RMSE metrics, they predict domain walls that differ by 40\%.

We also analyze these trends directly in configuration space in Fig~3e.
For UM2, the increased density of vectors in the AB region reflects the model's favouring of the AB stacking configuration. To compensate for its energy imbalance, UM2 also displays a sharper transition between stacking configurations, with a large density of disregistry vectors at the AA stacking configuration. Still, while these visual representations provide valuable insights, a quantitative comparison of the stacking disregistry distributions is still needed to rigorously assess the global structural discrepancies between models.

\subsection{Quantifying moiré structural dissimilarity: the mean disregistry error (MDE)}

We employ an approach rooted in optimal transport theory to quantify structural differences between moiré predictions from competing models. We adapt the concept of the Wasserstein distance -- a measure of the cost to transform continuous probability distributions into one another -- to our discrete case of disregistry vectors. Since disregistry vectors form a discrete set rather than a continuous distribution, we formulate this transformation as a minimum-cost bipartite matching problem, efficiently solvable via the Hungarian algorithm~\cite{kuhn1955hungarian}. This approach naturally connects to the physical concept of disregistry flow in configuration space during structural relaxation~\cite{carr2018relaxation} and captures \moire structural features that traditional RMSE metrics miss~\cite{sadeghi2013metrics}.

We hence propose a quantitative measure of the similarity between two moiré systems and, hence, the fitness of an MLIP: the \emph{mean disregistry error} (MDE), defined as the mean distance between two sets of matched disregistries vectors in configuration space. Critically, the MDE remains invariant under rigid translations, rotations, or atomic species relabeling, making it a robust metric for comparing moiré structures predicted by different models.

We can now assess whether the MDE correlates with conventional metrics for validating MLIPs, such as the energy RMSE. As training a comprehensive range of models with a uniform distribution of RMSEs is difficult, we instead apply controlled random noise to the weights of a single well-performing split MLIP model. We scale the noise relative to the standard deviation of the model weights within each neural network layer, thereby creating a continuous corruption factor. To minimize bias from the random noise, we average each calculated corruption level over an ensemble of ten initial random seeds. We analyze intralayer and interlayer models separately by fixing one component while systematically perturbing the other, allowing us to independently quantify the error contributions from the intralayer and interlayer MLIPs.

Our analysis (Fig. 3f) reveals a strong positive correlation between energy RMSE and moiré structural prediction quality as measured by MDE. However, we observe two important features that highlight the shortcomings of the RMSE as a validation metric for moiré materials. First, the RMSE metric exhibits significantly higher variance than the MDE (nearly half an order of magnitude) compared to the MDE for equivalent corruption factors. This is consistent with the idea that the MDE is more sensitive to domain symmetry and stacking distributions, making it less sensitive to fluctuation due to the model performance. Second -- and more critically --, for a fixed corruption factor and resultant MDE, the intralayer model energy RMSE is nearly two orders of magnitude larger than that for the interlayer model.

This finding has profound consequences for the design of MLIPs for moiré materials. It shows that one should train MLIPs with much a stricter RMSE for interlayer interactions than intralayer ones. Specifically, achieving an MDE below $0.01$ Å (comparable to the in-plane displacement associated with the zero-point motion of atoms in monolayer $\mos$) requires interlayer and intralayer energy RMSE values of approximately 4 meV/atom and 200 meV/atom, respectively. Given the impracticality of designing unified MLIPs that simultaneously satisfy such disparate accuracy requirements for moiré systems, these results provide quantitative justification for our split modelling approach that treats interlayer and intralayer interactions separately.

\subsection{Practical validation of MLIPs: surrogate quasi-1D structures}

While the MDE analysis is effective for assessing the accuracy of MLIPs on small systems, applying this framework to large-scale 2D moiré systems is computationally prohibitive due to the significant cost of generating high-accuracy, ground-truth relaxed structures for comparison. For instance, producing the long-range effects characteristic of moiré systems, such as domain reconstruction and out-of-plane buckling, requires simulation supercells with lattice constants exceeding $\sim$10 nm~\cite{weston2020atomic}. For twist angles below $1^\circ$, moiré lattices can surpass 20 nm, resulting in supercells with over 10,000 atoms~\cite{carr2017twistronics,carr2018relaxation,naik2018ultraflatbands} -- impractical for DFT calculations.

To address these challenges, we propose quasi-one-dimensional (Q1D) moiré structures as surrogate systems for validating MLIPs in 2D moiré materials. Q1D structures can be physically realized by uniaxially straining a bilayer structure and creating a supercell with a long moiré lattice parameter $\ell_m$ but a small width $w$. These structures retain the critical features of long-range moiré reconstruction along $\ell_m$ while reducing the number of atoms across their width~\cite{schleder2023one,cazeaux2017analysis}. This dimensionality reduction allows the number of atoms to scale linearly with $\ell_m$, in contrast to the quadratic scaling with \moire lattice constants for twisted 2D moiré systems. This scaling improvement enables one to directly perform ground-truth DFT calculations to assess the performance of MLIPs in capturing long-range moiré phenomena with the previously introduced MDE metric.

\begin{figure*}[t!]
    \centering
    \includegraphics[width=0.9\linewidth]{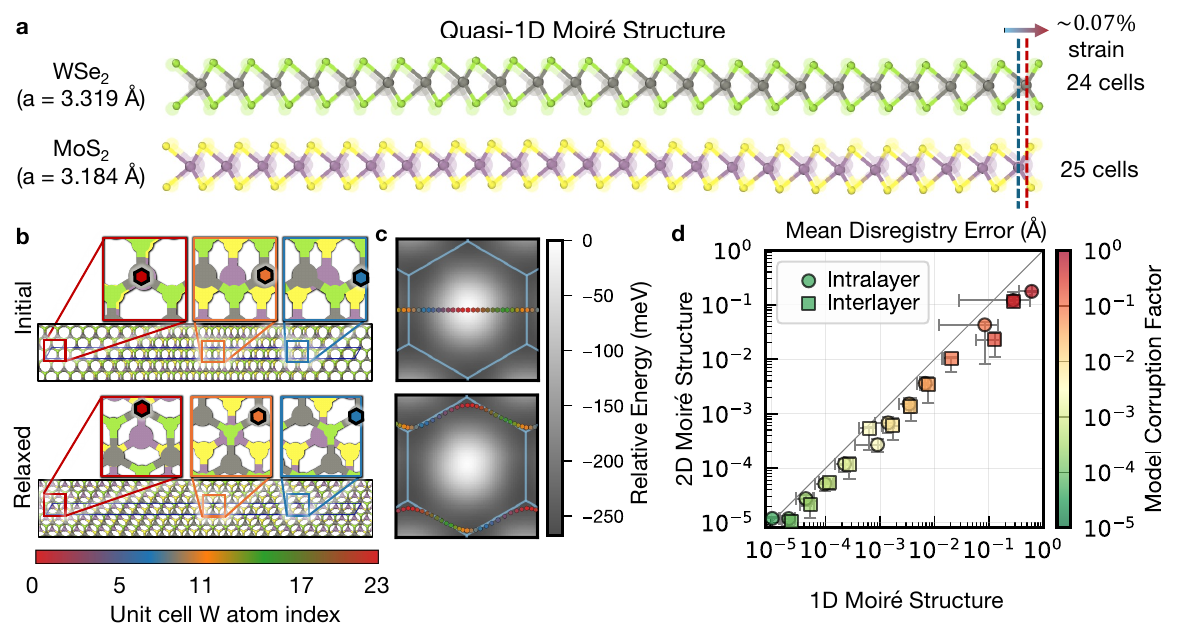}
    \caption{One-dimensional periodic moiré bilayers as a surrogate system for two-dimensional twisted structures.  \textbf{a} Quasi-1D (Q1D) heterostructure: 24-unit-cell strip of $\wse$ ($a=3.319$\AA) aligned with 25-unit-cell strip of $\mos$ ($a=3.184$\AA). The lattice mismatch requires $\sim$0.07\% strain in the $\wse$ layer to achieve periodicity. The initial, unrelaxed structure in transparency is overlaid onto the relaxed structure. Shifting of local disregistry and out-of-plane buckling is observed.  \textbf{b} Top view comparing unrelaxed (top) and DFT-relaxed (bottom) structures. Insets show local stacking configurations at three positions, with hexagons marking top-layer W atoms indicating relative disregistry from bottom-layer Mo atoms (coloured by indexed position along the strip). \textbf{c} Voronoi-centered disregistry analysis of unrelaxed (top) and relaxed (bottom) structures. Coloured hexagons represent disregistry vectors of W atoms, demonstrating relaxation-induced displacement along the short axis that reduces stacking energy while increasing strain.  \textbf{d} Validation of structural predictions, using the mean disregistry error (MDE),  comparing Q1D and 2D moiré structures. These MLIPs were trained against classical force fields to allow 2D \moire structures to be validated against a ground truth (see text). Models are corrupted following the same scheme as in Fig.~3, and error bars represent the 25th and 75th quartiles across multiple noise seed initializations. Notably, the performance of the split MLIP on the 1D structure is a good predictor of the quality of the MLIP on the 2D \moire structure.}
    \label{fig:1DP}
\end{figure*}

We present an illustrative Q1D structure comprising a stretched strip of 24 $\mathrm{\wse}$ unit cells on top of a compressed strip of 25 $\mathrm{\mos}$ unit cells (Fig 4a), generating a 1D moiré pattern associated with a $\sim0.7\%$ uniaxial strain (see SI~\cite{SI} for the detailed construction scheme).
Initially, the W and Mo atoms are aligned along the x-axis (Fig.~4b). Upon relaxation, the top layer shifts along the y-axis, producing a periodic modulation of stacking configurations along the length (Fig.~4c). This shift reflects the balance between stacking energies and elastic strain, as the system minimizes its total energy by favouring lower-energy stacking configurations (AB and BA) while limiting elastic deformation~\cite{carr2018relaxation}. In the initial, unrelaxed configuration, the Voronoi-centered disregistry vectors align along a linear path parallel to the x-axis. After relaxation, these vectors trace a sinusoid-like path near the Voronoi cell edges, reflecting the stacking modulation along the Q1D structure's length. This relaxation behaviour illustrates how the atoms migrate from a configuration path that passes through the high-energy near-AA stacking regions towards a path through the lower-energy AB and BA regions. We note that these Q1D systems could be generated along a diverse set of paths in configuration space to further strengthen the validation procedure.

To demonstrate the effectiveness of Q1D structures as surrogate systems to capture moiré relaxation in 2D twisted \moire systems, we compare the MDE predicted by MLIPs for both Q1D and 2D structures in Fig.~4d. We evaluate models on a 2D $\wse/\mos$ bilayer with a $2^{\circ}$ twist ($\sim$3,000 atoms) and the Q1D structure ($\sim$147 atoms). Because ground-truth DFT calculations with $\sim$3,000 atoms can be computationally expensive, we trained the split MLIPs here against classical force fields, which we then take as a surrogate for DFT calculations. This allows us to easily obtain \moire structures to be used in our ground-truth validation set.

We then use a controlled corruption factor applied to trained MLIPs, similar to that in Fig.~3f. Across all corruption levels, the MDEs between ground truth and predicted relaxations exhibit a strong positive correlation between Q1D and 2D moiré structures, with the Q1D structures displaying larger sensitivity to model accuracy (Fig.~4d). This strongly suggests that MLIPs performing well on Q1D structures will perform at least as well on the corresponding 2D ones, providing a conservative and reliable validation approach for MLIPs instead of the traditional, computationally intractable 2D \moire systems.

To summarize, we have presented an approach to systematically train and validate MLIPs for \moire reconstruction of large arbitrary bilayers. Further technical details on the practical usage of this approach are included in the SI~\cite{SI}. 

\subsection{Example: emergent triangular lattice in twisted bilayer $\vb*{\hfs}$/$\vb*{\gas}$} \label{sec:HfS2GaStext}

\begin{figure*}[htb!]
    \centering
    \includegraphics[width=1.0\linewidth]{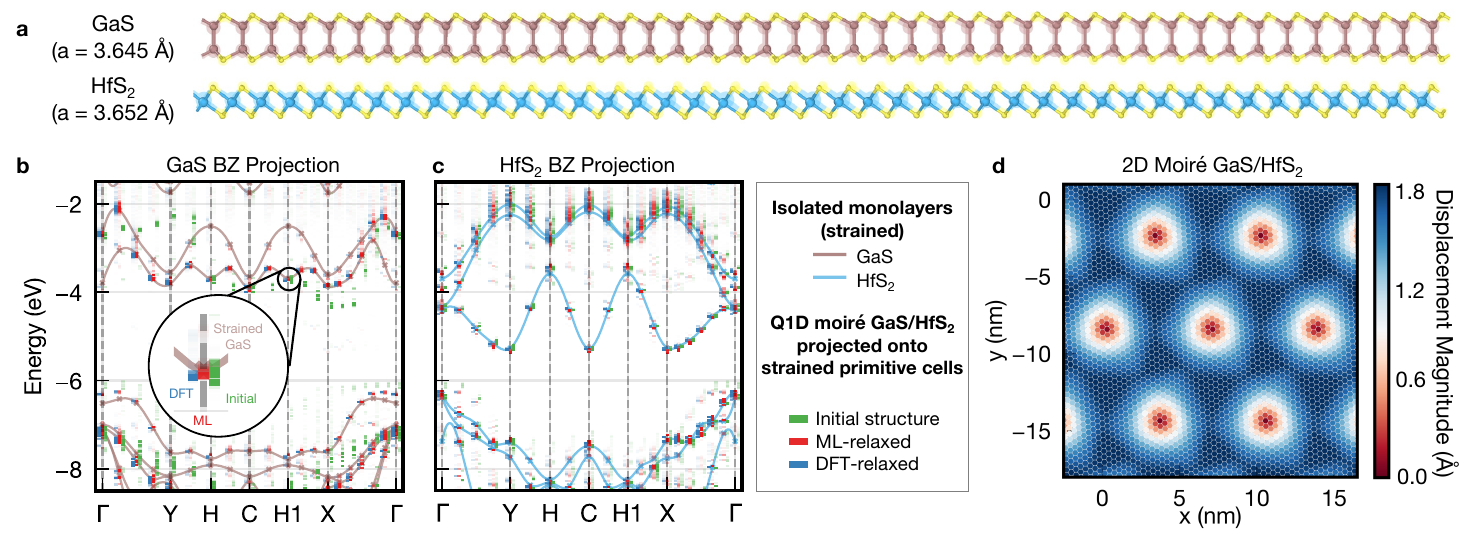}
    \caption{\textbf{a} Q1D structures of bilayer of HfS$_2$/GaS before (transparent circles) and after (solid circles) relaxation by our split MLIP model. \textbf{b} Band structure of the \Moire system unfolded onto a strained GaS unit cell (see SI Fig. 3~\cite{SI}), computed within DFT at the initial (green), MLIP-relaxed (red), and DFT-relaxed (blue) structures. Continuous curves denote the band structure of a single strained GaS monolayer. \textbf{c} Similar to \textbf{b}, but unfolded onto a strained HfS$_2$ cell. \textbf{d} MLIP-relaxed 2D \moire cell of GaS/HfS$_2$, along with the magnitude of the displacement vectors, showcasing the significant reconstruction and formation of triangular domains.}
    \label{fig:GaSHfS2}
\end{figure*}

To showcase the applicability of the surrogate approach to understudied 2D moiré bilayers, we apply it to predict the relaxations of the HfS$_2$/GaS bilayer.  HfS$_2$, known for its higher predicted mobility compared to traditional TMDs~\cite{zhang2014two}, has been studied as a promising material for 2D transistors~\cite{kanazawa2016few} and ultra-sensitive phototransistors~\cite{xu2015ultrasensitive}. Similarly, monolayer GaS, with its wider band gap, has been actively studied for its optical properties~\cite{gutierrez2022layered, chen2019ultrathin}. The comparable lattice parameters of these materials allow for the formation of large moiré lattices with potentially significant reconstruction. Recent first-principles calculations of HfS$_2$/GaS bilayers~\cite{wang2022first}, along with related gallium chalcogenide heterostructures~\cite{wang2018bandgap, obeid2020type, zou2021type}, reveal their potential for photocatalytic water splitting, attributed to a type-II band alignment and efficient charge separation, tunable via strain engineering. However, the relaxation of large moiré bilayers in this system has not been studied so far.

\begin{table}[htb!]
\centering
\caption{Validation of the split MLIP model for bilayer $\hfs$/$\gas$ on the quasi-1D structure. We compute several properties from DFT at either the MLIP-relaxed or the initial, rigidly strained structure and compare their deviation from the ground-truth values obtained at the DFT-relaxed structure. Comparison metrics include the DFT total energy error, the relative error in the Kohn-Sham DFT Hamiltonian, and the mean disregistry error (MDE).}
\begin{tabularx}{\columnwidth}
{|
>{\raggedright\arraybackslash}X|
>{\centering\arraybackslash}X|
>{\centering\arraybackslash}X|
>{\centering\arraybackslash}X|
}
\hline
 \makecell[l]{Structure} &
    \makecell[c]{DFT total\\energy error \\ (meV/atom)} &
    \makecell[c]{DFT\\Hamiltonian\\error (\%)} & \makecell[c]{MDE (Å)}\\
 \hline
  MLIP relaxed & 2.52 & 1.03 & 0.02 \\
  \hline
   Initial &  13.98 & 47.00 & 0.48 \\
 \hline
\end{tabularx}
\label{tab:quasi1D}
\end{table}

We first generate an interlayer dataset and train an interlayer model specifically for the HfS$_2$/GaS bilayer, while we model the intralayer interactions using a MACE foundational model. Using a generic, foundational model here is consistent with our finding that one can use a less accurate intralayer MLIP than compared to the intralayer one, and is sufficient, here, to capture the atomic and electronic structure for this bilayer. Following our Q1D approach, we construct a 1D commensurate cell (0.99° twist, 301 atoms, $\ell_m=$ 15.6 nm), shown inFig.~5a, and relax it using both DFT and our split MLIP framework. Comparing the two relaxed structures, we find a small MDE of 0.02 Å -- very close to our previously established, and somewhat arbitrary, desired threshold of 0.01 Å (Fig.~3f), despite employing a simplified intralayer MLIP from a foundational model. In contrast, the MDE between the initial structure and final DFT-relaxed structure is much larger, 0.48 Å, highlighting that the composite MLIP could capture the significant relaxation predicted from DFT (Table 1). Additionally, the energy of the MLIP-relaxed structure, when evaluated within DFT, is only 2.5 meV/atom higher than the DFT-relaxed structure – within the desired accuracy range.

Next, we assess the accuracy of the MLIP in predicting not only the moiré structure but also the electronic structure. We compute the electronic band structure, within DFT, of the Q1D structures relaxed with both DFT and the composite MLIP model. To make the interpretation easier, we unfold the computed band structures onto the primitive cell of either monolayer HfS$_2$ (Fig.~5b) or GaS (Fig.~5c), revealing that the electronic structure evaluated at the MLIP-relaxed structure is very close to one evaluated at the DFT-relaxed one. Finally, we quantify the similarity of the two electronic structures by computing the relative change of the occupied Kohn-Sham Hamiltonian, ${\lVert H  - H_{\mathrm{ref}}\rVert_{\mathrm{F}}}/{\lVert H_{\mathrm{ref}}\rVert_{\mathrm{F}}}$, where $H_{\mathrm{ref}}$ is the occupied subspace of the Kohn-Sham Hamiltonian obtained at the reference, DFT-relaxed structure, $H$ is the one evaluated at either the MLIP-relaxed or initial structure, and $\lVert \cdot \rVert_{\mathrm{F}}$ is the Frobenius norm (see SI section on computing error in the Hamiltonian~\cite{SI}). Notably, this difference is only 1\% for the MLIP-relaxed structure and 47\% for the initial structure. Altogether, this validation protocol using Q1D structures demonstrates the accuracy of the spit MLIP for HfS$_2$/GaS in accurately capturing atomistic relaxations necessary to describe moiré domains and the electronic structure in the system. 
 
Finally, we apply the MLIP to a relaxation of a 2D \moire superlattice (2.86° twist, 5163 atoms, $\ell_m=6.8$\, nm), revealing the formation of clear triangular-like domains in the relaxed structure (Fig.~5d). Such domains are qualitatively different from one obtained in twisted TMDs close to either 0° or 60° twist angles and hint at the possibility of realizing a Kagame lattice in such a system. Future DFT calculations utilizing large-scale, first-principles approaches or machine-learned Hamiltonians can resolve the emergent electronic properties of such bilayer systems now that the relaxed moiré structure -- which is challenging to obtain from DFT -- is accessible.

\section{Conclusion}
In this work, we established a comprehensive framework for constructing and rigorously validating MLIPs specifically designed for multilayer 2D materials. By explicitly separating the total energy and atomic forces into distinct intralayer and interlayer components, we demonstrated a tenfold improvement in predictive accuracy compared to conventional unified models, particularly highlighted in the $\mos$/$\wse$ material system. Our comparative analysis revealed significant limitations of traditional validation metrics, such as RMSE, which fails to capture the structural accuracy in large-scale moiré reconstructions.

To address these limitations, we introduced a physically intuitive and computationally efficient quantitative validation metric based on the Wasserstein distance, the mean disregistry error (MDE). This generalizable metric quantifies structural accuracy by comparing Voronoi-centered stacking disregistry vector distributions, thereby explicitly assessing the nm-scale distribution of local stacking disregistries crucial for moiré phenomena. By systematically evaluating models with controlled corruption, we established a clear correlation between energy RMSE and MDE metric, providing actionable accuracy thresholds for practical IP design: specifically, a stringent requirement of $\sim$ 4 meV/atom for interlayer interactions and more tolerant 200 meV/atom for intralayer interactions to achieve MDE values below $0.01$~Å. These findings provide a crucial, physically motivated justification for our split modelling approach. 

Furthermore, recognizing the computational impracticality of validating large two-dimensional moiré structures directly with first-principles methods, we proposed quasi-one-dimensional (Q1D) surrogate systems. These Q1D surrogates successfully preserve essential long-range structural and energetic features, making first-principles ground-truth validations computationally feasible. By demonstrating that MDE values in these surrogate systems correlate strongly with those in full 2D moiré structures, we confirmed Q1D systems as reliable validators for MLIPs. This allowed us to confidently extend validations to previously uninvestigated materials systems, exemplified by our analysis of the twisted HfS$_2$/GaS bilayer. Our method accurately captured subtle relaxation effects (achieving an MDE of 0.02~Å), predicted the formation of triangular domain patterns at small twist angles, and reproduced electronic structures consistent with high-fidelity DFT calculations, validating the robustness of our approach. 

Overall, the split interatomic approach developed here, combined with the quasi-1D surrogate validation strategy and the introduction of the MDE metric, provides a robust, scalable, and rigorously validation pathway for accurately predicting relaxation-driven emergent phenomena in these large-scale multilayer moiré systems. Importantly, this pipeline is generalizable and model-agnostic, readily adaptable to other layered material systems, and establishes a foundation for extending layered materials design beyond conventional twisted bilayer systems to more complex multilayer architectures with predictive reliability and computational efficiency.

\section{Methods}
\label{sec:methods}
\subsection{First-principles calculations}
\subsubsection{Dataset generation}
All DFT calculations for $\mos$ and $\wse$ were run using Quantum Espresso(v.7.2)~\cite{giannozzi_quantum_2009}. The self-consistent field (SCF) calculations were run with a plane-wave cutoff of $60$~Ry with a self-consistent field (SCF) convergence threshold of $10^{-8}$~Ry. An unshifted Monkhorst-Pack $\kvec$-point sampling of $12\times12\times1$ was used in the primitive cells and the structures in the interlayer dataset. For the intralayer dataset (constructed from a $3\times3\times1$ supercells of $\mos$ or $\wse$ primitive cells), an unshifted Monkhorst-Pack $\kvec$-point sampling of $4\times4\times1$ was used. For the calculations involving $\mos$ or $\wse$, we used the scaler-relativistic norm-conserving pseudopotentials from the PseudoDojo database~\cite{van2018pseudodojo}. For bilayer HfS$_2$/GaS, we use a plane-wave cutoff of $80$~Ry with an SCF convergence threshold of $10^{-6}$~Ry. For these calculations involving HfS$_2$ and GaS, we used the scaler-relativistic norm-conserving pseudopotentials from the SG15 database~\cite{schlipf2015optimization}. To include vdW interactions, all calculations use the \texttt{vdw-df-c09} functional~\cite{cooper2010van}. 

For the intralayer dataset, we generate 1,000 random perturbations to the atomic positions and cell parameters starting from a $3\times3$ supercell. 
These structures are then split randomly for training, validation, and test sets.

For the interlayer dataset, we first obtain an effective interlayer bonding distance, $l_0$, obtained by summing the covalent radii~\cite{cordero2008covalent} of the closest two atoms on different layers. We sample various interlayer separations $\ilspacing$ in steps of 0.1~Å between the bond length of the two closest atoms $l_0$ and 2$l_0$. We also sample $\ilspacing$ in steps of 0.5~Å between 2$l_0$ and 3$l_0$ to capture the decay of the long-range interaction. At every $\ilspacing$, we sample at different interlayer stacking registries by considering all possible in-plane layer displacements on a  $12\times12$ grid in fractional coordinates. To create a distinct test set, every second point from the $12\times12$ grid is set aside, forming a $6\times6$ test grid separate from the remaining $6\times6$ training and validation set. Overall, we generate 4,000 calculations for the interlayer dataset.

\subsubsection{Q1D relaxation and band structure of HfS$_2$/GaS}

DFT calculations for the quasi-one-dimensional (Q1D) HfS$_2$/GaS bilayer were run using Quantum Espresso(v.7.3)~\cite{giannozzi_quantum_2009}. The relaxation was performed using a BFGS optimizer from an initial rigidly twisted Q1D structure with a plane wave cutoff of $60$~Ry, force convergence threshold of $10^{3}$~Ry/Bohr, and a total energy convergence threshold of $10^{-4}$~Ry/Bohr. An unshifted Monkhorst-Pack $\kvec$-point sampling of $1\times12\times1$ was used, accounting for the supercell in the first lattice direction ($\mathbf{a}_1$) and the period, in-plane direction with the smallest lattice constant along the second direction ($\mathbf{a}_2$) of the Q1D structure. The SCF calculations for the initial, DFT-relaxed, and MLIP-relaxed structures were performed with a plane-wave cutoff of $60$~Ry with an SCF convergence threshold of $10^{-10}$~Ry on an unshifted Monkhorst-Pack $\kvec$-point sampling of $1\times12\times1$. The non-self-consistent field (NSCF) calculations for each structure were calculated on 20 $\kvec$-points along a path in the supercell Brillouin zone (BZ), which are equivalent to the high-symmetry $\kvec$-point paths with the individual strained HfS$_2$ and GaS monolayer BZ, up to multiples of the supercell reciprocal lattice vectors. We used the scaler-relativistic norm-conserving pseudopotentials from SG15 database~\cite{schlipf2015optimization}. 

To analyze the electronic structure in terms of the constituent primitive cells, we employed the primitive band unfolding method developed by Popescu and Zunger~\cite{popescu2012extracting}. Additionally, we resolve each Kohn-Sham (KS) according to the two layer projections. Specifically, we multiply each KS orbitals in the \moire system, in real-space, by a Heaviside function that is closer to either the HfS$_2$ or GaS layers.

\subsection{Interatomic potential parameterization and usage}
\subsubsection{Allegro IPs}
All Allegro MLIP models were trained using relatively standard hyperparameters and termination conditions (see Data Availability). For the interlayer models, We customize the Allegro model to include only interlayer interactions. Specifically, in the Allegro model~\cite{musaelian2023learning}, the total energy of the system is expressed as
\begin{equation}
    E = \sum_{j \in \neighbours(i)} \sigma_{Z_i, Z_j} E_{ij}
\end{equation}
where $Z_i$ is the atomic number of atom $i$, $E_{ij}$ is the pairwise energy and $\sigma_{Z_i, Z_j}$ is the per-species-pair scaling factor. We decompose the energy into its intralayer ($\Eintra$) and interlayer ($\Einter$) contributions,
\begin{equation}
\begin{split}
     \Eintra + \Einter =
        &\sum_{j \in \neighbours_{l_i}(i)} \sigma_{Z_i, Z_j} E_{ij} + \\
        &\sum_{j \in \neighbours(i), j \notin \neighbours_{l_i}(i)} \sigma_{Z_i, Z_j} E_{ij}.
\end{split}
\end{equation}

By using a distinct $Z$ for atoms in each layer, we can set $\sigma_{Z_i, Z_j}$ to 0 for the intralayer pairs. This modification procedure could be applied to other areas, such as defects, interfaces, or molecules on surfaces as well.
\subsubsection{Classical interatomic potentials}
All classical interatomic potentials were calculated in LAMMPS code~\cite{thompson2022lammps} using the pair styles available as \texttt{sw/mod} and \texttt{kolmogorov/crespi/z}, using previously reported parameters~\cite{naik2019kolmogorov}. Structure relaxation was performed using the FIRE optimization up to a force tolerance $10^{-4}$~eV/Å.

Corruption analyses were performed by introducing controlled random perturbations (corruption) into a single accurate split MLIPs model trained on classical interatomic potentials. A random Gaussian noise was added to the weights of the neural networks constituting the intralayer or interlayer MLIPs independently. The noise magnitude was scaled relative to the standard deviation of the original model parameters within each neural network layer, providing a controlled corruption factor. For each corruption level, we averaged results across ten different random seed initializations to increase statistical significance.

\section*{Acknowledgements}
The development of the split MLIP model was supported by the National Science Foundation (NSF) CAREER award through Grant No. DMR-2238328.
The validation metric and protocol were supported by the Office of Naval Research through the Multi-University Research Initiative (MURI) on Twist-Optics (Grant \# N00014-23-1-2567). This research used resources of the National Energy Research Scientific Computing Center (NERSC), a U.S. Department of Energy Office of Science User Facility located at Lawrence Berkeley National Laboratory, operated under Contract No. DE-AC02-05CH11231 using NERSC award BES-ERCAP ERCAP0024305 to generate the DFT datasets for $\mos$/$\wse$ and $\hfs$/$\gas$. The calculations of the 1D DFT relaxations and band structures of quasi-1D $\hfs$/$\gas$ used computational resources from the Texas Advanced Computing Center (TACC) at The University of Texas at Austin, funded by the National Science Foundation (NSF) award 1818253, through allocation DMR21077. J.D.G. acknowledges support from the Natural Science and Engineering Research Council (NSERC) of Canada through the Post-Graduate Scholarship PGS D-568202-2022.

\section*{Conflict of Interest}
The authors declare no conflict of interest.

\section*{Data Availablity Statement}
The data supporting the findings of this study are available in \url{https://github.com/jornada-group/mlips_split_data/}. We share the associated raw log information for our MLIP training in the \texttt{wandb} format~\cite{wandb}, available at \url{https://wandb.ai/twist-anything-ml/MLIPS_Paper?nw=mentxo4d3si}.  Any additional data that support the findings of this study are available from the corresponding author upon reasonable request.

\section*{Author Contributions}
F.H.J. conceived of the approach, led, and acquired funding for the research. J.D.G. and F.H.J. designed the structural similarity metric. J.D.G. performed the model and surrogate system validation. A.R. and F.H.J. developed the split model approach. A.R. wrote the training and dataset generation code and trained the models. J.D.G., A.R., and F.H.J. wrote the manuscript. C.H.S. and E.H. helped refine the dataset generation and model training processes. J.D.G and T.L. refined the surrogate system relaxation process. A.R. and B. extended the split MLIP approach to the MACE model.

\bibliographystyle{apsrev4-2}

\end{thebibliography}

\end{document}